\documentclass[conference]{IEEEtran}
\IEEEoverridecommandlockouts
\usepackage{amsmath,amssymb,amsfonts}
\usepackage{algorithmic}
\usepackage{graphicx}
\usepackage{textcomp}
\usepackage{xcolor}
\usepackage[backend=biber, style=ieee, maxnames=3, minnames=1]{biblatex}
\usepackage[hidelinks]{hyperref}
\usepackage[capitalize,noabbrev]{cleveref}
\usepackage{braket}
\usepackage{multirow}
\usepackage{booktabs}
\usepackage{pdflscape}
\usepackage{subfig}
\usepackage{tikz}

\graphicspath{ {./figures/} }

\def\BibTeX{{\rm B\kern-.05em{\sc i\kern-.025em b}\kern-.08em
    T\kern-.1667em\lower.7ex\hbox{E}\kern-.125emX}}

\newcommand\copyrighttext{%
\footnotesize \textcopyright 2025 IEEE. Personal use of this material is permitted.
Permission from IEEE must be obtained for all other uses, in any current or future
media, including reprinting/republishing this material for advertising or promotional purposes, creating new collective works, for resale or redistribution to servers or lists, or reuse of any copyrighted component of this work in other works. 
DOI: \href{https://doi.org/10.1109/QCE65121.2025.00268}{10.1109/QCE65121.2025.00268}
}
\newcommand\copyrightnotice{%
\begin{tikzpicture}[remember picture,overlay]
\node[anchor=south,yshift=10pt] at (current page.south) {\fbox{\parbox{\dimexpr\textwidth-\fboxsep-\fboxrule\relax}{\copyrighttext}}};
\end{tikzpicture}%
}

\begin{document}

\title{Quantum Autoencoder for Multivariate Time Series Anomaly Detection}

\author{
    \IEEEauthorblockN{
        Kilian Tscharke\IEEEauthorrefmark{1}\IEEEauthorrefmark{3}, Maximilian Wendlinger\IEEEauthorrefmark{1}, Afrae Ahouzi\IEEEauthorrefmark{1}, Pallavi Bhardwaj\IEEEauthorrefmark{2}, \\
        Kaweh Amoi-Taleghani\IEEEauthorrefmark{2}, Michael Schrödl-Baumann\IEEEauthorrefmark{2}, Pascal Debus\IEEEauthorrefmark{1}}
    \IEEEauthorblockA{
    \IEEEauthorrefmark{1}\textit{Fraunhofer Institute for Applied and Integrated Security (AISEC)}, Garching near Munich, Germany\\ 
    \IEEEauthorrefmark{2}\textit{SAP SE}, Walldorf, Germany\\
    \IEEEauthorrefmark{3}\{firstname.lastname\}@aisec.fraunhofer.de
    }
}

\maketitle
\copyrightnotice

\begin{abstract}
Anomaly Detection (AD) defines the task of identifying observations or events that deviate from typical -- or \textit{normal} -- patterns, a critical capability in IT security for recognizing incidents such as system misconfigurations, malware infections, or cyberattacks.
In enterprise environments like SAP HANA Cloud systems, this task often involves monitoring high-dimensional, multivariate time series (MTS) derived from telemetry and log data.
With the advent of quantum machine learning offering efficient calculations in high-dimensional latent spaces, many avenues open for dealing with such complex data. One approach is the Quantum Autoencoder (QAE), an emerging and promising method with potential for application in both data compression and AD. 
However, prior applications of QAEs to time series AD have been restricted to univariate data, limiting their relevance for real-world enterprise systems.

In this work, we introduce a novel QAE-based framework designed specifically for MTS AD towards enterprise scale. We theoretically develop and experimentally validate the architecture, demonstrating that our QAE achieves performance competitive with neural-network-based autoencoders while requiring fewer trainable parameters. We evaluate our model on datasets that closely reflect SAP system telemetry and show that the proposed QAE is a viable and efficient alternative for semisupervised AD in real-world enterprise settings.
\end{abstract}


\section{Introduction}
Anomaly Detection (AD) refers to the process of identifying patterns or events that deviate from typical -- or \textit{normal} -- behavior \cite{Ruff2020}. It plays a critical role in IT security and many other domains, as anomalies often correspond to potential security breaches, frauds, or system failures~\cite{zamanzadeh2023deep,wang2025survey}.
Modern enterprise infrastructure, such as SAP HANA Cloud and other large scale cloud native applications, rely on continuous monitoring to ensure optimal performance, availability, and reliability. With increasing system complexity and scale, observability platforms generate large volumes of telemetry data, including structured multivariate time series (MTS) and unstructured log streams. While traditional rule-based alerting systems are still widely used, they are inherently limited: they depend on manually defined thresholds, cannot adapt to complex dynamics, and often fail to detect subtle or novel anomalies. As infrastructure grows more distributed and dynamic, automated data-driven AD becomes increasingly vital for maintaining operational health.

A core challenge in enterprise observability, particularly in the context of SAP cloud systems, lies in effectively detecting anomalies in high-dimensional MTS data. Unlike univariate time series, MTS AD must account for temporal dependencies as well as correlations across variables, increasing modeling complexity~\cite{wang2025survey}. In real-world scenarios, labeled anomalies are rare or unavailable, necessitating semisupervised or unsupervised approaches that can learn from normal behavior alone. These challenges drive the need for scalable, accurate, and adaptive AD techniques for enterprise telemetry.

In recent years, deep learning-based approaches -- especially autoencoders (AEs) -- have shown promise in modeling complex system dynamics for AD. An AE is a neural network trained to reconstruct input data after compressing it into a latent representation. When trained on normal data only, it can effectively model the system’s baseline behavior. At inference time, inputs that the AE reconstructs poorly (i.e., with high error) are flagged as anomalous. This provides an unsupervised way to capture system dynamics and detect deviations without requiring labeled anomalies.

However, neural network-based AEs face limitations when applied to high-dimensional enterprise data. They often require large architectures with many parameters, which increases training time and may reduce generalization through overfitting. Quantum machine learning (QML) has emerged as a promising avenue to address these challenges by exploiting the unique capabilities of quantum computing. Quantum algorithms operate in Hilbert space with dimensionality that grows exponentially with the number of qubits, potentially enabling more compact representations and faster processing of complex, high-dimensional data than classical methods~\cite{biamonte2017quantum}. This potential has motivated explorations of quantum-enhanced AD, especially as NISQ quantum hardware becomes available.

Among QML approaches, the Quantum Autoencoder (QAE) is particularly well suited for semisupervised AD, making it a promising candidate for enterprise health monitoring. In this work, we apply QAEs to public MTS datasets that closely resemble telemetry data from SAP HANA Cloud systems. The model assigns anomaly scores to input windows, enabling the detection of irregular patterns without requiring labeled anomalies during training. This study marks an initial step toward integrating QML into enterprise observability pipelines by introducing a novel industrial application of QML and exploring its feasibility for real-time AD in high-dimensional MTS data. To evaluate the proposed QAE, we compare its performance against classical AE baselines using six standard metrics: AUC, precision, recall, F1-score, accuracy, and balanced accuracy.

The remainder of this work is structured as follows:
The next \cref{Related Work} offers a comprehensive review of research in the field of QAEs and their application to AD.
Our contributions are detailed in \cref{Contributions}.
The following Preliminaries (\cref{Preliminaries}) presents the fundamentals required to understand this work, specifically anomalies, MTS, the window-based approach to AD, and neural-network AEs.
Next, the Proposal (\cref{Proposal}) introduces our QAE method for MTS AD.
Afterwards, the Experiments (\cref{Experiments}) describe the implementation of the (Q)AE and give an overview of the datasets used for the benchmark.
In the Results and Discussion (\cref{Results and Discussion}), we analyze the performance of our QAE and the results of the benchmark.
Finally, Conclusion and Outlook (\cref{Conclusion and Outlook}) highlights the key results of this work and provides future research directions.

\subsection{Related Work} \label{Related Work}
The QAE was originally proposed by Romero et al.~\cite{romero2017quantum} in 2017, and mimics classical AEs using quantum circuits to encode and reconstruct quantum states. They applied the model to compress ground states of the Hubbard model and molecular Hamiltonians.
Kottmann et al.~\cite{kottmann2021variational} extended this concept to AD by showing that QAEs could identify quantum phase transitions. Their QAE relied solely on a variational quantum eigensolver to prepare the ground states and an encoder to decouple a subset of the qubits -- the \emph{trash qubits} -- from the rest of the system, effectively compressing the original ground state into a smaller number of qubits. The trash qubits were then measured and used to calculate a compression metric, which simultaneously served as an anomaly score. 
The training of the model is done by a classical feedback loop, where the calculation of the loss is the only part performed on a quantum computer.
They demonstrated the feasibility of generating the phase diagram of the one-dimensional extended Bose–Hubbard model with dimerized hoppings using a single training sample.

Recently, QAEs have been explored for AD across a variety of domains, demonstrating the versatility of QAEs in handling AD tasks under diverse and often imbalanced conditions. Applications include detecting anomalies at the Large Hadron Collider (LHC)~\cite{QAE_high_energy_physics, QAE_LHC}, identifying network intrusions~\cite{QAE_network_security}, and flagging fraudulent credit card transactions~\cite{QAE_credit_card}. In addition, QAEs have been employed as quantum baselines in comparative studies evaluating the performance of various QML models for AD tasks~\cite{QAE_tscharke}.

Finally, Frehner and Stockinger~\cite{frehner2024applying} applied QAEs to classical time series data, demonstrating that they can outperform neural-network AEs in AD tasks while requiring orders of magnitude fewer trainable parameters. They further validated their approach on real quantum hardware, showing that the QAE maintains its AD performance despite hardware-induced noise. However, their study was limited to univariate time series and relied only on subsets of the full datasets. The extension of QAEs towards MTS data of industrially-relevant size remains an open challenge -- one that this work aims to address.

\subsection{Contributions} \label{Contributions}
In this paper, we propose the first QAE architecture tailored for MTS AD. Our main contributions are as follows:
\begin{itemize}
  \item We design a parameter-efficient, problem-agnostic QAE that jointly encodes multivariate temporal features, enabling the model to capture inter-variable dependencies.
  \item We demonstrate that our QAE achieves competitive performance on two MTS and one univariate dataset. Notably, the \emph{SMD} dataset closely reflects telemetry data from SAP systems, underscoring the model's potential for real-world deployment in enterprise health monitoring and observability pipelines.
  \item We demonstrate that our approach requires significantly fewer trainable parameters than a reasonably-sized neural-network AE, while maintaining comparable performance.
\end{itemize}

Overall, our work demonstrates that QAEs can be effectively extended to multivariate settings, paving the way for quantum-enhanced AD in real-world scenarios such as IT infrastructure monitoring, industrial control systems, and cyber-physical networks.

\section{Preliminaries} \label{Preliminaries}
In the following, the fundamentals required to understand the proposal of this work are presented. First, a definition of an anomaly is given, and different types of anomalies are described. Next, MTS is formally defined before explaining the window-based AD approach. Finally, the general structure of an AE is given, setting the basis for our proposed QAE architecture.

\subsection{Anomalies}
Anomalies are patterns or observations that deviate from the expected behavior of a system \cite{Ruff2020}. These deviations may signal critical events such as faults, intrusions, or system failures, making their detection essential in various application domains such as IT security. Depending on how they manifest in the data, anomalies can be broadly categorized into three types: point anomalies, contextual anomalies, and collective anomalies.

\subsubsection{Point Anomalies}
Point anomalies are individual observations that deviate significantly either from their immediate neighbors (local outliers) or from the overall distribution of the dataset (global outliers). For example, an unexpected spike in network traffic during off-peak hours may be flagged as a point anomaly. Such anomalies typically indicate isolated, unexpected events.

\subsubsection{Contextual Anomalies}
Contextual anomalies arise when a data point appears normal in a global sense but is anomalous within a specific context, which is often defined by temporal or spatial attributes. For instance, high CPU usage in data centers may be expected during business hours or periods of high website traffic, however, such usage outside these time windows or in the absence of corresponding traffic may indicate anomalous behavior. Detecting contextual anomalies requires an understanding of both the data and the context in which it occurs.

\subsubsection{Collective Anomalies}
Collective anomalies involve a sequence or group of data points that together form an anomalous pattern, even if the individual points appear to be normal. For example, a prolonged period of low CPU usage on a server that normally handles high loads might suggest a malfunction, although each individual reading could fall within the normal range. Detecting such patterns typically requires models that can capture temporal dynamics and inter-variable dependencies, making MTS analysis particularly well-suited for this type of anomaly.

\subsection{Multivariate Time Series}
A \textit{time series} is an ordered set of observations indexed by time, typically collected at regular intervals \cite{blazquez2021review}. In a \textit{univariate} time series, each observation $x_t \in \mathbb{R}$ for $0 \leq t \leq T$ is a scalar, forming a sequence $X = (x_0, x_1, \dots, x_T)$. In contrast, a \textit{multivariate} time series consists of $d$-dimensional vectors at each time step, denoted as $X = (\mathbf{x}_0, \mathbf{x}_1, \dots, \mathbf{x}_T)$ with $\mathbf{x}_t \in \mathbb{R}^d$. The latter formulation enables the modeling of both temporal dependencies and inter-variable relationships, which are particularly important in applications such as system monitoring, finance, and IT security \cite{wang2025survey}. Thus, anomalies can manifest in MTS even if each variable considered individually seems normal on its own, due to potentially  anomalous interactions among them.

\subsection{Window-Based Anomaly Detection}
While point anomalies can often be trivially detected using just rule-based approaches based on mean and standard deviation, contextual anomalies require, as the name implies, sufficient contextual information for detection. A common and effective strategy for detecting collective anomalies in MTS is to partition the data into potentially overlapping \textit{windows} of fixed length $L$ and stride $S$. Each window $W_i$ is defined as
\begin{equation}
    W_i = (\mathbf{x}_{i,0}, \mathbf{x}_{i,1}, \dots, \mathbf{x}_{i,L-1}),
\end{equation}
for $i = 0, 1, \dots, N$, where $N = \frac{T - L}{S} + 1$. Each $W_i \in \mathbb{R}^{L \times d}$ thus represents a segment of $L$ consecutive time steps, capturing the temporal dynamics and multivariate structure of the input sequence. For multivariate AD, all time steps of a window $i$ are concatenated into a single flattened vector $\mathbf{w}_i$, which serves as the input to the AD models.

In the \textit{window-wise labeling} paradigm, each window is associated with a binary label $y_i \in \{0, 1\}$, where $y_i = 1$ indicates the presence of at least one anomalous observation within the window. 
While this type of preprocessing keeps the temporal order within each window intact, it can also be interpreted as reframing AD from a sequence-level task to a point-level task, where each point is now a higher-dimensional point corresponding to a full window. This enables the use of a broader range of machine learning techniques beyond those specialized in sequences.

\subsection{Autoencoders} \label{Autoencoders}
The following section gives a brief definition of AEs, a neural-network-based approach suited for semi- and unsupervised learning. A more in-depth explanation is given in \cite{vincent2008extracting, Goodfellow-et-al-2016}

\begin{figure}[htbp]
\centerline{\includegraphics[width=.5\textwidth]{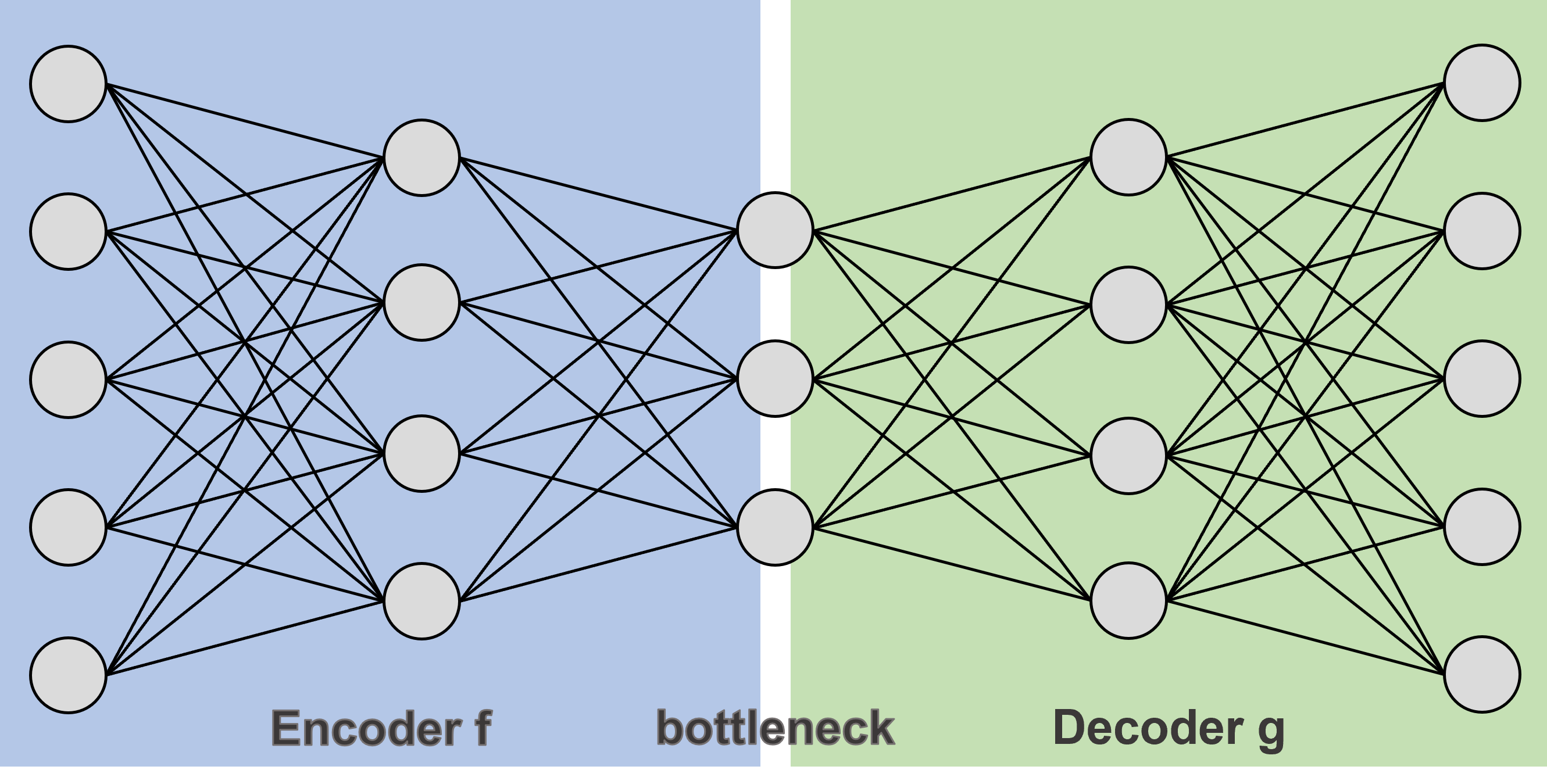}}
\caption{General architecture of an AE consisting of encoder and decoder with the bottleneck layer.}
\label{fig:autoencoder_architecture}
\end{figure}

The AE consists of an encoder and a decoder, as shown in \cref{fig:autoencoder_architecture}.
Formally, let $\mathbf{x} \in \mathbb{R}^d$ represent an input vector, e.g., a flattened MTS window. The encoder function $f_{\theta}(\cdot)$ maps $\mathbf{x}$ to a latent representation $\mathbf{h} \in \mathbb{R}^p$ as follows:
\begin{equation}
    \mathbf{h} = f_{\theta}(\mathbf{x}) = s(\mathbf{W}\mathbf{x} + \mathbf{b}),
\end{equation}
where $\mathbf{W} \in \mathbb{R}^{p \times d}$ is the weight matrix, $\mathbf{b} \in \mathbb{R}^p$ is the bias vector, and $s(\cdot)$ denotes a nonlinear activation function such as the sigmoid or ReLU.
In most cases, the latent dimension is chosen to be smaller than the input dimension, i.e., $p < d$, resulting in what is commonly referred to as the \emph{information bottleneck}. This design enforces a compression of the input data, encouraging the AE to learn a compact representation that captures the most salient features of the original input.

Subsequently, the decoder function $g_{\theta'}(\cdot)$ maps the latent representation $\mathbf{h}$ back to the reconstructed input $\mathbf{x'} \in \mathbb{R}^d$: 
\begin{equation} 
\mathbf{x'} = g_{\theta'}(\mathbf{h}) = s'(\mathbf{W'}\mathbf{h} + \mathbf{b'}), 
\end{equation} 
where $\mathbf{W'} \in \mathbb{R}^{d \times p}$ and $\mathbf{b'} \in \mathbb{R}^d$ are the decoder's weight matrix and bias vector, respectively, and $s'(\cdot)$ is the decoder's activation function.

The training objective is to minimize a \emph{reconstruction error} or loss that quantifies the difference between the input $\mathbf{x}$ and its reconstruction $\mathbf{x'}$. A common choice for this loss function is the mean squared error (MSE): 
\begin{equation}
    \mathcal{L}(\mathbf{x}, \mathbf{x'}) = \|\mathbf{x} - \mathbf{x'}\|^2.
\end{equation}
By minimizing this error, the AE learns to capture the underlying structure of normal data. Since anomalies typically deviate from this learned distribution, they are reconstructed with higher error, making the loss a natural candidate for an anomaly score.

\section{Quantum Autoencoder for Multivariate Time Series Anomaly Detection} \label{Proposal}
This section introduces the QAE architecture for MTS AD, drawing upon the framework proposed by Kottmann et al. \cite{kottmann2021variational}.
Unlike classical AEs, which consist of separate encoder and decoder networks, the QAE for AD does not require an explicit decoder.
Instead, after encoding the input data into a parameterized quantum circuit, a designated subset of qubits -- the  \emph{trash qubits} -- is measured, and the result is used to compute the loss function.
The core principle of this approach is to train the variational parameters of the quantum circuit such that, for normal data, the trash qubits become disentangled from the rest of the system and collapse to the $\ket{1}$-state. This behavior indicates that the essential information of the input data has been compressed onto the remaining unmeasured qubits, and the trash qubits do not carry information about the input as their state is independent from it. Once the QAE is trained, it can effectively detect anomalies since anomalous data cannot be compressed as efficiently, and hence the trash qubits cannot be decoupled from the rest of the system. For anomalous data, the trash qubits deviate from the $\ket{1}$-state,  resulting in an increased loss.

\subsection{Quantum Autoencoder Architecture}
The QAE is an $n$-qubit parametrized quantum circuit consisting of a series of variational and entangling gates (defined as a general unitary $U$) followed by a measurement of the trash qubits $S$. The output of the model $f$ is the sum of the probabilities of the trash qubits being in the $\ket{0}$-state given by
\begin{equation} \label{eq:model}
f(\boldsymbol{x} ; \boldsymbol{w}, \boldsymbol{b}) = \frac{1}{2}\sum_{s\in S} \left( 1 + \braket{Z_s}\right)
\end{equation}
where $\braket{Z_s}$ is the $Z$-expectation value of trash qubit $s$ after applying $U$, i.e. 
\begin{equation} \label{eq:observable}
\braket{Z_s} = \left\langle 0\left|U(\boldsymbol{x} ; \boldsymbol{w}, \boldsymbol{b})^{\dagger} Z_s U(\boldsymbol{x} ; \boldsymbol{w}, \boldsymbol{b})\right| 0\right\rangle .
\end{equation}
Here, $U(\boldsymbol{x} ; \boldsymbol{w}, \boldsymbol{b})$ is the parametrized unitary that encodes both the data $\boldsymbol{x}$ and the trainable parameters (weights $\boldsymbol{w}$ and bias $\boldsymbol{b}$ with $\boldsymbol{w}, \boldsymbol{b} \in \mathbb{R}^{L\times n\times3}$) and $\mathbf{x}$ is a single datum vector (a flattened window for MTS data). 

This circuit employs \emph{data reuplod-encoding} \cite{reuploading_encoding}, an encoding strategy where the input data is sequentially encoded multiple times throughout the circuit. Such an approach enhances the expressivity of the quantum model, allowing it to approximate a broader class of functions. Theoretical analyses have shown that repeated data encoding interleaved with trainable operations can expand the accessible frequency spectrum of the model, thereby increasing its capacity to represent intricate functions \cite{Schuld_data_encoding}. This makes data re-uploading a valuable technique in designing expressive QML models.

Our unitary for this encoding strategy consists of $L$ layers of general single-qubit rotation gates followed by a sequence of CNOT gates forming an entangling block. The rotation gate $R$ is parametrized by three angles, allowing the realization of arbitrary single-qubit rotations \cite{Quantum_Computing_Schuld2021}. Commonly, $R$ is defined and decomposed into three base rotations as
\begin{align}
R(\phi, \theta, \omega) &= 
\begin{bmatrix}
e^{-i(\phi + \omega)/2} \cos(\theta/2) & -e^{i(\phi - \omega)/2} \sin(\theta/2) \notag \\
e^{-i(\phi - \omega)/2} \sin(\theta/2) & e^{i(\phi + \omega)/2} \cos(\theta/2)
\end{bmatrix} \\
&= RZ(\omega)\, RY(\theta)\, RZ(\phi) 
\end{align}
Paired with the trainable parameters $\boldsymbol{w}$ and $\boldsymbol{b}$, we can thus write the individual unitaries used in the QAE as 
\begin{equation} \label{eq:rot-gate}
    U_j(\boldsymbol{x};\ \boldsymbol{w}_j, \boldsymbol{b}_j) = R\left(\boldsymbol{w}_j \circ \boldsymbol{x} + \boldsymbol{b}_j \right).
\end{equation}
As detailed, these unitaries are stacked into $L$ layers, including an entangling block at the end of each layer to allow inter-qubit dependencies via entanglement. Note the subtle difference between individual unitaries $U_j$ used inside the circuit's layers and the general unitary $U$ defining the whole state progress of the zero states as a result of circuit operations.

\begin{figure}[htb] 
\centering
\includegraphics[width=0.49\textwidth]{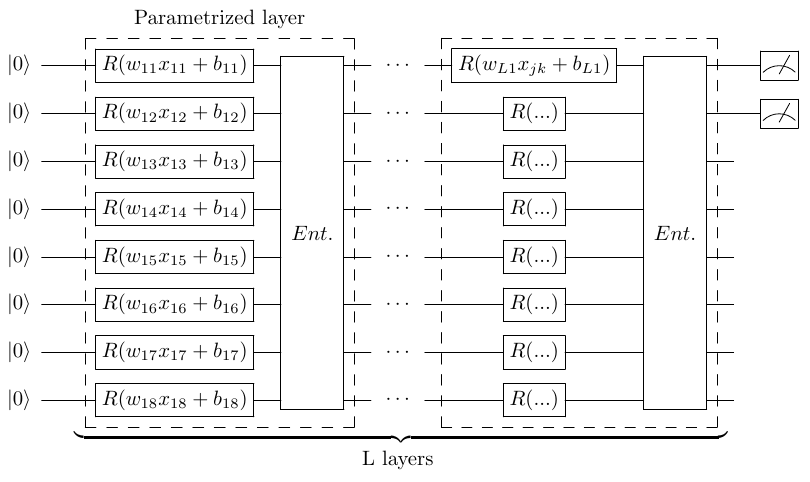}
\caption{The QAE is realized using a trainable re-upload encoding architecture including multiplicative weight and additive bias parameters. Detailed explanation in text.}
\label{fig:reup-circ}
\end{figure}

An exemplary circuit consisting of 8 qubits and 2 trash qubits, including the parametrized layers and measurement, is shown in \cref{fig:reup-circ}. Each $x_{jk}$ is a 3-tuple of features of the input vector $\mathbf{x}$, i.e. $x_{11} = (x_0, x_1, x_2)$ and $x_{12} = (x_3, x_4, x_5)$ and so on.
The input is encoded multiple times if $3*L* n > d$.

\subsection{Model Training and Inference}
In the semi-supervised setting, the model is trained solely on data assumed to be normal. The model's output itself is used as a loss function, delivering minimal loss if the trash qubits are in $\ket{1}$:
\begin{equation}
\mathcal{L} =  f(\boldsymbol{x} ; \boldsymbol{w}, \boldsymbol{b}),
\end{equation}
During inference, a sample is classified as \textit{normal} if the output loss is below a predefined threshold; otherwise, it is labeled as \textit{anomalous}.
Unlike in the original formulation by Kottmann et al.~\cite{kottmann2021variational}, where the loss is minimized when the trash qubits are in the $\ket{0}$ state, we define our loss to be minimized when the trash qubits are in the $\ket{1}$ state. This modification improves trainability under our reupload-encoding strategy, which differs from the encoding employed in \cite{kottmann2021variational}.

To illustrate the importance of this change, consider the alternative loss function minimized when the trash qubits occupy the $\ket{0}$ state:
\begin{equation} \label{eq:wrong_model}
f(\boldsymbol{x} ; \boldsymbol{w}, \boldsymbol{b}) = \frac{1}{2} \sum_{s \in S} \left(1 - \braket{Z_s} \right),
\end{equation}
For this loss function, the model exhibits a trivial global minimum. By setting all weights and biases to zero, the output state remains invariant for any input:
\begin{align} \label{eq:unitary_identity}
U_l(\boldsymbol{x}; \boldsymbol{w} = 0, \boldsymbol{b} = 0) \ket{0}^{\otimes n} 
&= \text{CNOT} \cdot R^{\otimes n}(0 \cdot \boldsymbol{x} + 0) \ket{0}^{\otimes n} \notag \\
&= \text{CNOT} \ket{0}^{\otimes n} = \ket{0}^{\otimes n},
\end{align}
for any layer $U_l$ composed of single-qubit rotations $R$ and any arbitrary sequence of CNOTs. As a result, all trash qubits remain in the $\ket{0}$ state, leading to $\braket{Z_s} = 1$ for all $s \in S$, and consequently zero loss according to \cref{eq:wrong_model}.

Although this solution perfectly classifies normal data, it fails to capture meaningful distinctions in the input space and thus cannot detect anomalies. Even if the quantum circuit starts from a nontrivial initial state, the model can exploit the re-uploading encoding by using the bias terms to map the state back to $\ket{0}^{\otimes n}$, while keeping all weights zero.

In contrast, our modified loss function requires the model to map the trash qubits to the $\ket{1}$ state. Achieving this configuration is nontrivial due to the presence of CNOT gates, which introduce entanglement as soon as a single qubit deviates from $\ket{0}$. Therefore, the model cannot simply default to an identity operation or a fixed pattern. When combined with regularization of weights and biases, our formulation avoids trivial solutions and encourages the model to learn nontrivial, data-dependent representations -- thus enabling effective AD.

\section{Experiments} \label{Experiments}
\subsection{Quantum Autoencoder}
We use the same QAE architecture and hyperparameters for all datasets. The hyperparameters of the model were optimized using the \textit{machine-1-1} of SMD and are summarized in \cref{tab:qae_params}. 
Training and evaluation were performed on a statevector simulator specifically optimized for efficient batch processing \cite{quack}.
To improve the trainability of the model, the weights and biases are initialized with values close to zero \cite{BPs_init}. While the simulator is deterministic, the initialization of weights and biases introduces a probabilistic component. However, due to their near-zero initialization, the model is expected to exhibit low sensitivity to these initial values.
Anomalies were defined using the 99th percentile of the training reconstruction error.

\begin{table}[htbp]
\centering
\caption{Hyperparameters of QAE.}\label{tab:qae_params}
\begin{tabular}{ll}
\toprule
Parameter & Value \\
\midrule
qubits & 8 \\
layers & 100 \\
measure qubits & [0,1] \\
reg param weights & $10^{-2}$ \\
reg param bias & $10^{-4}$ \\
torch seed & 42 \\
scaler & MinMaxScaler \\
clipping & [0,1] \\
epochs & 200 \\
batch size & 32 \\
learning rate & $10^{-3}$ \\
early stop threshold & $10^{-5}$ \\
patience & 10 \\
threshold & $99^{\text{th}}$ percentile of train error \\
\bottomrule
\end{tabular}

\end{table}

\subsection{Datasets}
The MTS datasets SMD and Pasta, and the univariate time series dataset MSCM used for the benchmark of the QAE and their preprocessing are now described in more detail.
\cref{tab:datasets_overview} lists the statistics of the datasets.
After splitting the data into train and test sets, the anomalous windows in the train set are removed. 

\begin{table*}[htbp]
\centering
\caption{Overview of the datasets used for benchmarking.}\label{tab:datasets_overview}
\begin{tabular}{ll|rrr|rr|rrr}
\hline
 &  & \multicolumn{3}{c|}{Statistics} & \multicolumn{2}{c|}{Preprocessing} & \multicolumn{3}{c}{Splits} \\
\cline{3-10}
 &  & Timestamps & Dim. & Gran. & Win & Stride & \% Test An. & \# Train & \# Test \\
Dataset & Subset &  &  &  &  &  &  &  &  \\
\hline
\multirow[c]{13}{*}{MSCM \cite{microsoft_cloud_monitoring_dataset}} & api-01 & 6192 & 1 & 1h & 10 & 5 & 0.05 & 705 & 246 \\
 & app1-01 & 358 & 1 & 1h & 10 & 5 & 0.62 & 39 & 13 \\
 & app1-02 & 710 & 1 & 1h & 10 & 5 & 0.74 & 84 & 27 \\
 & app1-04 & 705 & 1 & 1h & 10 & 5 & 0.36 & 33 & 28 \\
 & app1-05 & 699 & 1 & 1h & 10 & 5 & 0.38 & 64 & 29 \\
 & app1-06 & 684 & 1 & 1h & 10 & 5 & 0.13 & 48 & 31 \\
 & app1-08 & 710 & 1 & 1h & 10 & 5 & 0.48 & 74 & 27 \\
 & app2-04 & 1118 & 1 & 1h & 10 & 5 & 0.07 & 133 & 44 \\
 & app2-05 & 1118 & 1 & 1h & 10 & 5 & 0.32 & 98 & 44 \\
 & app2-06 & 1116 & 1 & 1h & 10 & 5 & 0.09 & 128 & 43 \\
 & app2-07 & 1109 & 1 & 1h & 10 & 5 & 0.16 & 88 & 43 \\
 & ingress-02 & 15840 & 1 & 1m & 10 & 5 & 0.02 & 1900 & 632 \\
 & machine-01 & 20160 & 1 & 1m & 10 & 5 & 0.02 & 2363 & 805 \\
\hline
\multirow[c]{3}{*}{Pasta \cite{pasta_dataset}} & B1 & 1798 & 42 & 1d & 10 & 5 & 0.98 & 51 & 89 \\
 & B3 & 1798 & 21 & 1d & 10 & 5 & 0.65 & 109 & 89 \\
 & B4 & 1798 & 10 & 1d & 10 & 5 & 0.40 & 100 & 89 \\
\hline
\multirow[c]{8}{*}{SMD \cite{server_machine_dataset}} & machine-1-1 & 56959 & 5 & - & 100 & 50 & 0.12 & 568 & 568 \\
 & machine-1-2 & 47389 & 5 & - & 100 & 50 & 0.07 & 472 & 472 \\
 & machine-1-3 & 47406 & 5 & - & 100 & 50 & 0.08 & 473 & 473 \\
 & machine-1-4 & 47414 & 5 & - & 100 & 50 & 0.08 & 473 & 473 \\
 & machine-1-5 & 47412 & 5 & - & 100 & 50 & 0.03 & 473 & 473 \\
 & machine-1-6 & 47378 & 5 & - & 100 & 50 & 0.28 & 472 & 472 \\
 & machine-1-7 & 47395 & 5 & - & 100 & 50 & 0.15 & 472 & 472 \\
 & machine-1-8 & 47398 & 5 & - & 100 & 50 & 0.11 & 472 & 473 \\
\hline
\end{tabular}
\end{table*}

\subsubsection{Server Machine Dataset}
The Server Machine Dataset (SMD) \cite{server_machine_dataset} is a large-scale MTS dataset collected over five weeks from 28 different machines grouped into three categories within a production datacenter of a major internet company. Its high temporal resolution, substantial duration, and multivariate structure make it a strong benchmark for evaluating semisupervised AD models in realistic, industrial settings. Notably, the dataset closely reflects the telemetry characteristics of SAP systems, making it particularly relevant to our use case. Given the focus of this work on MTS-based AD, SMD serves as a primary dataset for empirical evaluation.

Each machine in SMD constitutes a separate subset, and models are trained and evaluated independently on each. The dataset is partitioned equally into a training set and a test set. Anomalies are present only in the test set, and labels are provided at each timestamp to indicate whether a point is anomalous. Additionally, SMD includes interpretation labels that identify which dimensions contributed to each detected anomaly -- offering fine-grained diagnostic information beyond binary classification.

In our experiments, we selected the following five features that contribute to most of the anomalies according to the interpretation labels for all machines: \texttt{load\_1}, \texttt{disk\_r}, \texttt{disk\_svc}, \texttt{disk\_w}, and \texttt{disk\_wb}. While the interpretation of these feature names is not properly documented in the original publication, in a general cloud telemetry context, those features most likely refer to the "Average number of processes waiting for CPU execution over 1 minute", "Number of Disk Read Operations", "Average time to complete a disk I/O request", "Number of Disk Write Operations", and "Disk Write Bytes", respectively.  With a window size of 100 time steps and 5 features per time step, this results in an input dimensionality of 500. Observations are sampled at one-minute intervals and are evenly spaced in time. No separate validation set is used, as the hyperparameter tuning is performed solely for \emph{machine-1-1}.

\subsection{Pasta Dataset}
The Pasta dataset~\cite{pasta_dataset} consists of real-world daily sales data collected from an Italian grocery store between January 1, 2014, and December 31, 2018. It includes 118 univariate time series representing the demand for various pasta products. In addition to the quantity sold, each data point indicates whether a promotion was active, although no further details on the promotion type or discount level are available.
The time series follows a natural hierarchical structure across three levels. At the top (Level~0) is the fully aggregated store-level demand series. This is disaggregated into four brand-level series (B1 to B4) at Level~1. Each brand-level series is further subdivided at Level~2 into individual item-level time series, corresponding to specific pasta products. The bottom-level series contains 42, 45, 21, and 10 items, respectively, for B1 through B4, capturing fine-grained sales dynamics across different pasta types and brands.

For this work, we combine the individual products within each brand into a single MTS, where the number of dimensions equals the number of products of that brand. A time point is labeled as anomalous if at least one product from the brand is under promotion at that time. After segmenting the data into fixed-size windows and applying window-wise labeling, this approach results in a high proportion of anomalous windows, as shown in \cref{tab:datasets_overview}. For brand B2, all windows in test sets were labeled as anomalous, rendering the subset unsuitable for evaluation. Consequently, B2 was excluded from the experiments.

\subsubsection{Microsoft Cloud Monitoring Dataset}
The Microsoft Cloud Monitoring (MSCM) \cite{microsoft_cloud_monitoring_dataset} dataset is a collection of real-world univariate time series derived from production telemetry signals across Microsoft services and clients. It was developed to support the design, evaluation, and improvement of AD algorithms used in Microsoft's internal cloud monitoring tools.
The dataset comprises 67 time series from eight application domains, including API query rates, database latencies, crash rates, and usage statistics. Each time series contains minute- or hour-level granular observations and is labeled at the timestamp level to indicate the presence of anomalies. Anomalies were identified and annotated by domain experts using a dedicated labeling tool, with some time series intentionally containing no anomalies to test false positive robustness.
As the dataset consists of univariate series with low inter-series correlation, it is not suitable for evaluating MTS AD models. Nevertheless, it provides a valuable benchmark for assessing detection accuracy on diverse, real-world, production-scale telemetry signals.
After partitioning the datasets into training and test sets and segmenting them into windows, we excluded all subsets that did not contain any anomalies in the test splits. This filtering step ensures meaningful evaluation of AD performance. As a result, our experiments are conducted on the 13 subsets listed in \cref{tab:datasets_overview}. The trained model on \emph{ingress-02} was corrupted, hence this subset could not be used for the benchmark.

\subsection{Neural-Network Autoencoder}
For benchmarking purposes, three neural-network-based AEs of varying sizes are used as baselines. Specifically, the AEs are configured with hidden layer architectures of [3], [16, 8], and [256, 128], respectively, and employ the \textit{ReLU} activation function. All other hyperparameters are kept consistent with those of the QAE in \cref{tab:qae_params} to ensure a fair comparison.

The motivation for evaluating AEs of different sizes lies in comparing models with varying representational capacities. For the SMD dataset, the smallest AE ([3]) has a comparable number of trainable parameters to the QAE, while the medium ([16, 8]) and large ([256, 128]) AEs have significantly more. This design allows us to investigate whether performance improvements stem from architectural differences or simply increased model capacity. A detailed comparison of the number of trainable parameters across datasets is provided in \cref{tab:num_trainable_params}.

\begin{table}[htbp]
\centering
\caption{Number of trainable parameters of the models.}\label{tab:num_trainable_params}
\begin{tabular}{lrrrr}
\toprule
Dataset & QAE & AE [3] & AE [16, 8] & AE [256, 128] \\
\midrule
MSCM & 2 400 & 60 & 288 & 70 656 \\
Pasta B1 & 2 400 & 2 520 & 13 696 & 280 576 \\
Pasta B3 & 2 400 & 1 260 & 6 976 & 173 056 \\
Pasta B4 & 2 400 & 600 & 3 456 & 116 736 \\
SMD & 2 400 & 3 000 & 16 256 & 321 536 \\
\bottomrule
\end{tabular}

\end{table}

\section{Results and Discussion} \label{Results and Discussion}
We evaluate four models -- our proposed QAE and three classical neural network-based AEs with increasing model capacity -- across two MTS datasets and one univariate time series dataset, and six performance metrics: AUC, precision, recall, F1-score, accuracy, and balanced accuracy.

\subsection{SMD Dataset}
The results on the large-scale MTS dataset SMD, which closely reflects telemetry data from SAP systems, are presented in \cref{tab:SMD} in \cref{app:detailed_results} in the appendix. 
Overall, the performance of the models shows that the SMD dataset is challenging for both our QAE and neural-network-based AEs.
Our proposed model, \textbf{QAE}, performs competitively with the classical baselines, achieving the \textbf{second-highest mean accuracy (0.74)} across all subsets. 
Furthermore, the QAE model achieves higher mean balanced accuracy than the small AE and medium AE models (0.67 vs. 0.63 and 0.64), despite using fewer trainable parameters.
QAE demonstrates consistently strong performance across several SMD subsets. On \textit{m-1-6}, it achieves the highest values in all evaluation metrics except recall, where it still performs well with a score of 0.98. It also attains the \textbf{highest accuracy} on subsets \textit{m-1-4}, \textit{m-1-6}, and \textit{m-1-7}, and maintains an accuracy above 0.80 on most other subsets, except for \textit{m-1-5}, where all models struggle. 
These reliable, good accuracies highlight its ability to detect anomalies in a semi-supervised setting. This advantage becomes particularly relevant when the decision threshold is selected based solely on the training data -- a common and realistic constraint in semisupervised AD scenarios. In such cases, the QAE demonstrates good robustness and generalization, effectively distinguishing between normal and anomalous patterns without relying on access to labeled anomalies before the inference stage.

\begin{figure*}%
    \centering
    \subfloat[\centering machine-1-1]{{\includegraphics[width=0.46\textwidth]{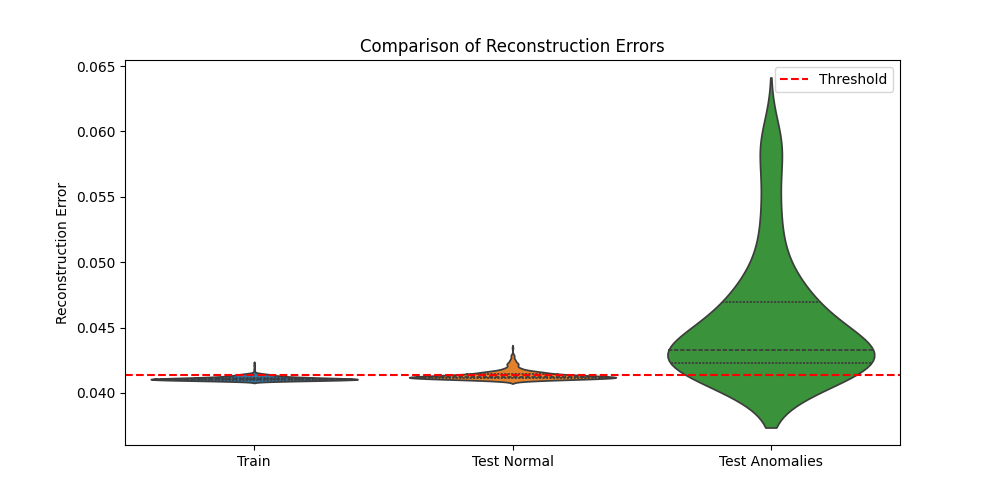} }}%
    \qquad
    \subfloat[{\centering machine-1-5}]{{\includegraphics[width=0.46\textwidth]{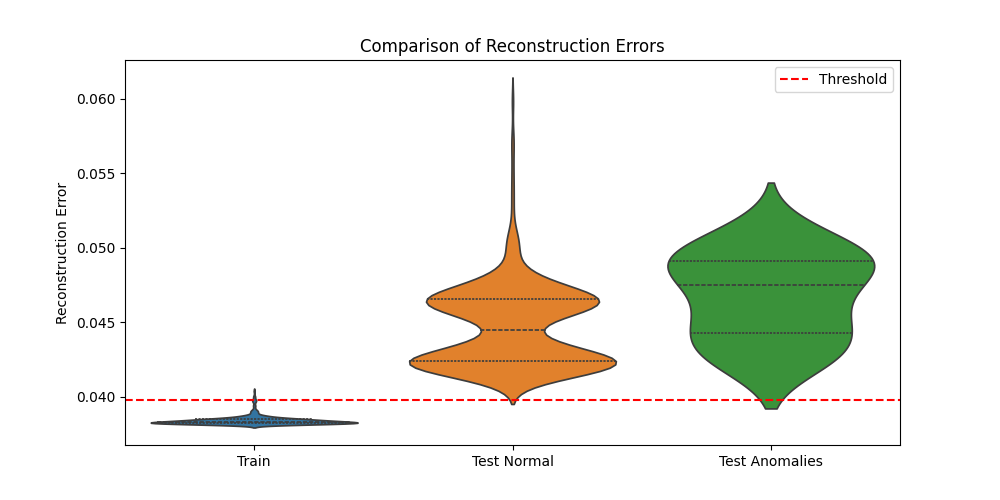} }}%
    \caption{Violin plots illustrating the distribution of reconstruction errors of QAE on two different machines of the SMD dataset. }%
    \label{fig:QAE_rec_loss}%
\end{figure*}

The reconstruction loss distributions of QAE for two representative subsets are shown in \cref{fig:QAE_rec_loss}. For \textit{m-1-1}, the reconstruction errors of anomalous test samples are clearly higher than those of the normal test samples. This is evident from the median of the anomalies (dashed line), which aligns with the upper extreme of the reconstruction errors for normal data. Furthermore, the lower quartile of the anomalous distribution lies above the upper quartile of the normal distribution, indicating a strong separation between the two classes. This clear distinction is reflected in the high AUC and accuracy achieved by QAE on this subset.
In contrast, the distribution of reconstruction errors on \textit{m-1-5} reveals substantial overlap between normal and anomalous test samples. Additionally, the reconstruction loss for the training data is significantly lower than that of the normal test samples, and the resulting threshold -- determined from training data -- is positioned near the lower end of both test distributions. As a consequence, the model struggles to separate the two classes effectively.
These observations suggest that QAE is overfitting on \textit{m-1-5}, learning to reconstruct the training data too well while failing to generalize to unseen normal and anomalous samples. This lack of generalization reduces the separability of the classes and results in the lower AUC and accuracy observed for this subset.

\subsection{Pasta Dataset}
The results on the Pasta dataset are presented in \cref{tab:pasta} in \cref{app:detailed_results} in the appendix. In contrast to SMD, the Pasta dataset features fewer subsets and smaller sample sizes but introduces challenges due to the higher dimensionality of the time series and high class imbalance, particularly for \textit{B1}.
Our proposed model, \textbf{QAE}, performs competitively with the classical baselines, achieving the \textbf{highest mean accuracy (0.72)} among all models, and achieving a mean AUC only 0.01 worse than that of the large AE, despite the QAE having drastically fewer trainable parameters, as shown in \cref{tab:num_trainable_params}.
The mean performance of the QAE is superior to the one of the small AE in all metrics but precision and comparable to the one of the medium AE, which has more trainable parameters than our QAE. 
Considering individual subsets, QAE achieves the best AUC on \textit{B1} and only 0.01 less AUC than the large AE on \emph{B2}. In comparison, the large AE obtains the highest mean AUC (0.79), F1-score (0.80), and recall (1.00), but at the cost of lower balanced accuracy and less consistent performance across subsets. AE [3] and AE [16, 8] show strong precision but generally lower recall, which leads to lower F1-scores and balanced accuracies overall.
These results indicate that QAE performs comparably to the AEs, particularly for the semisupervised setting when the threshold is selected based solely on the training data.

\begin{figure*}%
    \centering
    \subfloat[\centering QAE]{{\includegraphics[width=0.46\textwidth]{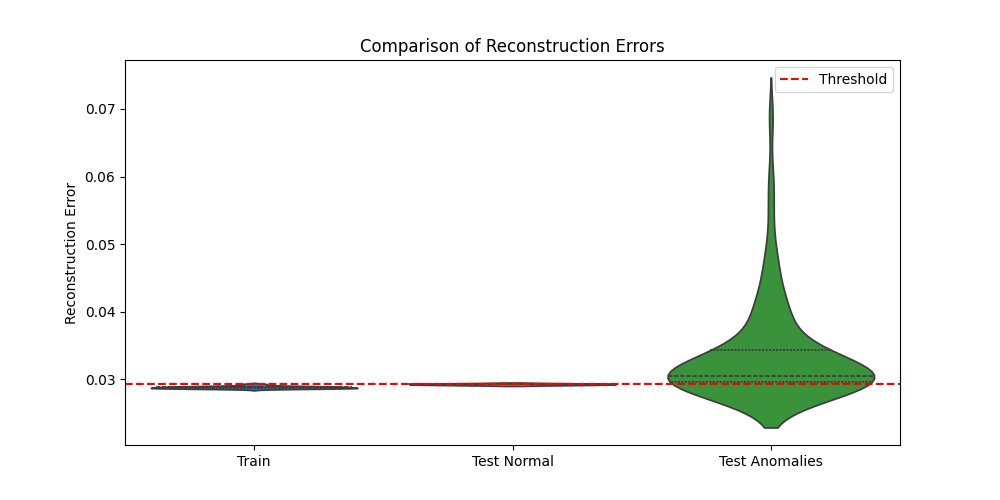} }}%
    \qquad
    \subfloat[{\centering AE [256, 128]}]{{\includegraphics[width=0.46\textwidth]{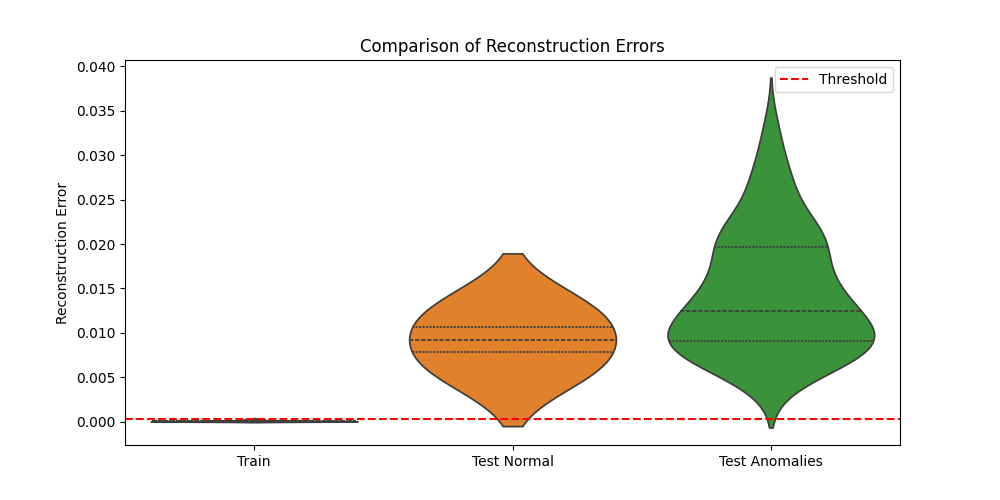} }}%
    \caption{Reconstruction errors of QAE and the large AE on B1 of the Pasta dataset. }%
    \label{fig:pasta_rec_loss}%
\end{figure*}

The reconstruction loss distributions of the QAE and the large AE on subset \textit{B1} are shown in \cref{fig:pasta_rec_loss}. It is important to note that the test set for this subset is highly imbalanced, containing approximately 98\% anomalous samples. For the QAE, the majority of anomalous test samples exhibit reconstruction errors above the threshold. However, around one quarter of the anomalies fall below the threshold, while the few normal test samples are located near the threshold. As a result, the model struggles to separate the classes perfectly, but still achieves a moderate balanced accuracy of 0.67.
In contrast, the large AE fails to distinguish between normal and anomalous samples. The reconstruction errors of both classes lie mostly above the threshold, and their distributions overlap substantially. This complete lack of separability prevents effective classification and results in a balanced accuracy of 0.50 -- equivalent to random guessing.

\subsection{MSCM Dataset}
The results on the MSCM dataset are presented in \cref{tab:MSCM} in \cref{app:detailed_results} in the appendix. As a diverse collection of real-world univariate time series from production cloud systems, MSCM provides a benchmark for evaluating AD models across a range of signal types and anomaly sparsity levels.

All models demonstrate strong overall performance, with mean AUC values exceeding 0.89. The proposed \textbf{QAE} achieves a \textbf{mean AUC} of \textbf{0.91}, matching the performance of the medium-sized AE and outperforming both the small and large AEs in this metric. When compared to the large AE -- which contains significantly more trainable parameters -- QAE achieves better performance across all metrics except recall, highlighting its efficiency and competitiveness.

In terms of mean accuracy, QAE ranks second (0.87), just behind the medium AE (0.89). It shows particularly strong results on the subsets \textit{app1-01}, \textit{app1-02}, \textit{app1-05}, \textit{app2-05}, \textit{app2-06}, \textit{app2-07}, and \textit{m-01}, where it achieves the highest accuracy and F1-score among all models.

Overall, QAE performs on the same level as the classical AEs, and the relatively weak performance of the large AE shows that parameter count is not the most important factor for model performance for this dataset.

\section{Conclusion and Outlook} \label{Conclusion and Outlook}
In this work, we proposed a QAE architecture tailored for AD in MTS, motivated by the practical challenges of monitoring enterprise systems such as SAP HANA Cloud. We extended the QAE concept to handle MTS input, enabling it to model complex temporal and inter-variable dependencies using a parameter-efficient quantum circuit design.

Our experimental results demonstrate that the performance of the QAE on MTS data -- while not ideal -- can compete with that of neural-network AEs, even when using significantly fewer trainable parameters. In particular, the QAE performs well when the decision threshold must be derived solely from normal training data -- an increasingly common requirement in enterprise monitoring pipelines.
Furthermore, our QAE employs a fixed architecture and identical hyperparameters across all MTS and univariate datasets, highlighting the problem-agnostic nature of the approach. This design choice underscores its potential for deployment across diverse domains without the need for task-specific tuning of the circuit ansatz or hyperparameters.

While promising, our results also reveal challenges such as overfitting or moderate performance on certain subsets. This limitation suggests that improvements are needed in circuit design or regularization strategies to promote better generalization across varied data distributions.

As a first step toward integrating quantum models into enterprise observability pipelines, our approach opens the door to several promising directions. Future work includes scaling the architecture to larger MTS inputs, deploying it on real quantum hardware, and investigating approaches to mitigate overfitting and improve training dynamics. With continued progress in QML and quantum hardware availability, QAEs may eventually offer a practical, resource-efficient alternative for real-time AD in complex industrial systems.

\section*{Acknowledgment}
SAP would like to acknowledge the Federal Ministry for Economic Affairs and Climate Action (abbreviated BMWK), for funding this work under the QCHALLenge Project (01MQ22008).
Fraunhofer AISEC would like to acknowledge the Munich Quantum Valley, which is supported by the Bavarian state government with funds from the Hightech Agenda Bayern Plus.

\printbibliography

\onecolumn
\appendices
\section{Detailed Results} \label{app:detailed_results}

\begin{table*}[htbp]
\centering
\caption{Results for the SMD dataset.}\label{tab:SMD}
\begin{tabular}{lllllllllll}
\toprule
 & Dataset & \multicolumn{9}{c}{SMD} \\
\cline{2-11}
 & Subset & 1-1 & 1-2 & 1-3 & 1-4 & 1-5 & 1-6 & 1-7 & 1-8 & \textbf{mean} \\
\midrule
\multirow[c]{6}{*}{QAE} & AUC & 0.97 & 0.62 & 0.84 & 0.86 & 0.72 & \textbf{0.74} & 0.75 & 0.65 & 0.77 \\
 & Precision & 0.38 & 0.31 & 0.73 & 0.71 & 0.03 & \textbf{0.34} & 0.87 & 0.44 & 0.48 \\
 & Recall & 0.96 & 0.15 & 0.49 & 0.71 & \textbf{1.00} & 0.98 & \textbf{0.36} & 0.15 & 0.60 \\
 & F1 & 0.55 & 0.20 & 0.58 & \textbf{0.71} & 0.06 & \textbf{0.51} & \textbf{0.50} & 0.23 & 0.42 \\
 & Acc. & 0.81 & 0.92 & 0.94 & \textbf{0.95} & 0.03 & \textbf{0.46} & \textbf{0.89} & 0.89 & 0.74 \\
 & Bal. Acc. & 0.87 & 0.56 & 0.74 & 0.84 & 0.50 & \textbf{0.62} & \textbf{0.67} & 0.57 & 0.67 \\
\cline{1-11}
\multirow[c]{6}{*}{AE [3]} & AUC & 0.88 & \textbf{0.87} & 0.84 & \textbf{0.91} & 0.77 & 0.72 & 0.78 & 0.51 & 0.79 \\
 & Precision & 0.43 & \textbf{1.00} & 0.89 & \textbf{1.00} & \textbf{0.05} & 0.29 & \textbf{1.00} & 0.24 & 0.61 \\
 & Recall & 0.59 & 0.21 & 0.41 & 0.37 & 0.87 & 0.99 & 0.08 & 0.17 & 0.46 \\
 & F1 & 0.49 & 0.35 & 0.56 & 0.54 & \textbf{0.10} & 0.45 & 0.15 & 0.20 & 0.36 \\
 & Acc. & \textbf{0.86} & 0.94 & 0.95 & 0.95 & \textbf{0.52} & 0.32 & 0.86 & 0.85 & \textbf{0.78} \\
 & Bal. Acc. & 0.74 & 0.61 & 0.70 & 0.68 & \textbf{0.69} & 0.52 & 0.54 & 0.55 & 0.63 \\
\cline{1-11}
\multirow[c]{6}{*}{AE [16, 8]} & AUC & 0.96 & 0.86 & \textbf{0.88} & 0.90 & \textbf{0.86} & 0.67 & 0.81 & 0.70 & 0.83 \\
 & Precision & \textbf{0.45} & 0.68 & \textbf{1.00} & 1.00 & 0.03 & 0.33 & 1.00 & \textbf{0.72} & \textbf{0.65} \\
 & Recall & 0.99 & 0.52 & 0.08 & 0.21 & 1.00 & 0.88 & 0.19 & 0.25 & 0.51 \\
 & F1 & \textbf{0.62} & \textbf{0.59} & 0.14 & 0.35 & 0.06 & 0.48 & 0.32 & 0.37 & 0.37 \\
 & Acc. & 0.85 & \textbf{0.95} & 0.92 & 0.94 & 0.03 & 0.46 & 0.88 & \textbf{0.91} & 0.74 \\
 & Bal. Acc. & \textbf{0.91} & \textbf{0.75} & 0.54 & 0.61 & 0.50 & 0.59 & 0.60 & 0.62 & 0.64 \\
\cline{1-11}
\multirow[c]{6}{*}{AE [256, 128]} & AUC & \textbf{0.98} & 0.86 & 0.87 & 0.89 & 0.84 & 0.73 & \textbf{0.82} & \textbf{0.76} & \textbf{0.84} \\
 & Precision & 0.28 & 0.12 & 0.81 & 0.50 & 0.03 & 0.28 & 0.92 & 0.29 & 0.40 \\
 & Recall & \textbf{1.00} & \textbf{0.94} & \textbf{0.67} & \textbf{0.82} & 1.00 & \textbf{1.00} & 0.33 & \textbf{0.62} & \textbf{0.80} \\
 & F1 & 0.44 & 0.21 & \textbf{0.73} & 0.62 & 0.06 & 0.44 & 0.48 & \textbf{0.39} & \textbf{0.42} \\
 & Acc. & 0.69 & 0.51 & \textbf{0.96} & 0.92 & 0.03 & 0.29 & 0.89 & 0.79 & 0.64 \\
 & Bal. Acc. & 0.82 & 0.71 & \textbf{0.83} & \textbf{0.87} & 0.50 & 0.50 & 0.66 & \textbf{0.71} & \textbf{0.70} \\
\bottomrule
\end{tabular}
\end{table*}

\begin{table*}[htbp]
\centering
\caption{Results for the Pasta dataset.}\label{tab:pasta}
\begin{tabular}{llllll}
\toprule
 & Dataset & \multicolumn{4}{c}{Pasta} \\
\cline{2-6}
 & Subset & B1 & B3 & B4 & \textbf{mean} \\
\midrule
\multirow[c]{6}{*}{QAE} & AUC & \textbf{0.85} & 0.71 & 0.76 & 0.78 \\
 & Precision & 0.99 & 0.89 & 0.68 & 0.85 \\
 & Recall & 0.84 & 0.41 & 0.72 & 0.66 \\
 & F1 & 0.91 & 0.56 & 0.70 & 0.72 \\
 & Acc. & 0.83 & 0.58 & 0.75 & \textbf{0.72} \\
 & Bal. Acc. & 0.67 & \textbf{0.66} & 0.75 & 0.69 \\
\cline{1-6}
\multirow[c]{6}{*}{AE [3]} & AUC & 0.71 & 0.55 & 0.77 & 0.68 \\
 & Precision & 0.98 & \textbf{1.00} & 0.75 & 0.91 \\
 & Recall & 0.54 & 0.26 & 0.67 & 0.49 \\
 & F1 & 0.70 & 0.41 & 0.71 & 0.60 \\
 & Acc. & 0.54 & 0.52 & 0.78 & 0.61 \\
 & Bal. Acc. & 0.52 & 0.63 & 0.76 & 0.64 \\
\cline{1-6}
\multirow[c]{6}{*}{AE [16, 8]} & AUC & 0.76 & 0.67 & 0.90 & 0.78 \\
 & Precision & \textbf{1.00} & 1.00 & \textbf{0.96} & \textbf{0.99} \\
 & Recall & 0.57 & 0.28 & 0.67 & 0.51 \\
 & F1 & 0.73 & 0.43 & \textbf{0.79} & 0.65 \\
 & Acc. & 0.58 & 0.53 & \textbf{0.85} & 0.66 \\
 & Bal. Acc. & \textbf{0.79} & 0.64 & \textbf{0.82} & \textbf{0.75} \\
\cline{1-6}
\multirow[c]{6}{*}{AE [256, 128]} & AUC & 0.75 & \textbf{0.72} & \textbf{0.92} & \textbf{0.79} \\
 & Precision & 0.98 & 0.65 & 0.44 & 0.69 \\
 & Recall & \textbf{1.00} & \textbf{1.00} & \textbf{1.00} & \textbf{1.00} \\
 & F1 & \textbf{0.99} & \textbf{0.79} & 0.62 & \textbf{0.80} \\
 & Acc. & \textbf{0.98} & \textbf{0.65} & 0.49 & 0.71 \\
 & Bal. Acc. & 0.50 & 0.50 & 0.58 & 0.53 \\
\bottomrule
\end{tabular}
\end{table*}

\begin{table*}[htbp]
\centering
\caption{Results for the MSCM dataset.}\label{tab:MSCM}
\setlength{\tabcolsep}{3pt}
\begin{tabular}{lllllllllllllll}
\toprule
 & Dataset & \multicolumn{13}{c}{MSCM} \\
\cline{2-15}
 & Subset & api-01 & app1-01 & app1-02 & app1-04 & app1-05 & app1-06 & app1-08 & app2-04 & app2-05 & app2-06 & app2-07 & m-01 & \textbf{mean} \\
\midrule
\multirow[c]{6}{*}{QAE} & AUC & 0.90 & 0.88 & \textbf{0.90} & 0.82 & \textbf{0.90} & \textbf{1.00} & 0.94 & 0.94 & 0.99 & 0.86 & 0.83 & 0.95 & \textbf{0.91} \\
 & Precision & 0.60 & \textbf{0.78} & \textbf{0.86} & 0.47 & \textbf{1.00} & 0.36 & 0.77 & 0.33 & \textbf{1.00} & \textbf{1.00} & \textbf{1.00} & \textbf{0.52} & 0.72 \\
 & Recall & 0.25 & 0.88 & \textbf{0.95} & 0.80 & \textbf{0.82} & \textbf{1.00} & 0.77 & 0.67 & \textbf{0.93} & \textbf{0.75} & \textbf{0.71} & 0.88 & 0.78 \\
 & F1 & 0.35 & \textbf{0.82} & \textbf{0.90} & 0.59 & \textbf{0.90} & 0.53 & 0.77 & 0.44 & \textbf{0.96} & \textbf{0.86} & \textbf{0.83} & \textbf{0.65} & 0.72 \\
 & Acc. & 0.96 & \textbf{0.77} & \textbf{0.85} & 0.61 & \textbf{0.93} & 0.77 & 0.78 & 0.89 & \textbf{0.98} & \textbf{0.98} & \textbf{0.95} & \textbf{0.98} & 0.87 \\
 & Bal. Acc. & 0.62 & \textbf{0.74} & \textbf{0.76} & 0.65 & \textbf{0.91} & 0.87 & 0.78 & 0.78 & \textbf{0.96} & \textbf{0.88} & \textbf{0.86} & 0.93 & 0.81 \\
\cline{1-15}
\multirow[c]{6}{*}{AE [3]} & AUC & 0.89 & 0.90 & 0.71 & 0.81 & 0.90 & 1.00 & 0.95 & 0.95 & \textbf{0.99} & 0.81 & 0.84 & 0.96 & 0.89 \\
 & Precision & \textbf{0.75} & 0.78 & 0.82 & 0.57 & 1.00 & 0.67 & \textbf{1.00} & \textbf{0.50} & 1.00 & 1.00 & 0.83 & 0.44 & 0.78 \\
 & Recall & 0.25 & 0.88 & 0.70 & 0.80 & 0.82 & 1.00 & 0.69 & 0.67 & 0.93 & 0.50 & 0.71 & \textbf{0.94} & 0.74 \\
 & F1 & 0.38 & 0.82 & 0.76 & 0.67 & 0.90 & 0.80 & \textbf{0.82} & \textbf{0.57} & 0.96 & 0.67 & 0.77 & 0.60 & 0.73 \\
 & Acc. & 0.96 & 0.77 & 0.67 & 0.71 & 0.93 & 0.94 & \textbf{0.85} & \textbf{0.93} & 0.98 & 0.95 & 0.93 & 0.97 & 0.88 \\
 & Bal. Acc. & 0.62 & 0.74 & 0.64 & 0.73 & 0.91 & 0.96 & \textbf{0.85} & \textbf{0.81} & 0.96 & 0.75 & 0.84 & \textbf{0.96} & 0.81 \\
\cline{1-15}
\multirow[c]{6}{*}{AE [16, 8]} & AUC & 0.91 & \textbf{0.95} & 0.66 & 0.89 & 0.89 & 1.00 & \textbf{0.95} & 0.96 & 0.96 & \textbf{0.90} & \textbf{0.86} & 0.95 & 0.91 \\
 & Precision & 0.62 & 0.78 & 0.82 & 0.56 & 1.00 & \textbf{0.80} & 1.00 & 0.40 & 1.00 & 1.00 & 1.00 & 0.44 & \textbf{0.79} \\
 & Recall & \textbf{0.83} & 0.88 & 0.70 & \textbf{1.00} & 0.82 & 1.00 & 0.69 & 0.67 & 0.93 & 0.50 & 0.71 & 0.94 & 0.81 \\
 & F1 & \textbf{0.71} & 0.82 & 0.76 & 0.71 & 0.90 & \textbf{0.89} & 0.82 & 0.50 & 0.96 & 0.67 & 0.83 & 0.60 & \textbf{0.77} \\
 & Acc. & \textbf{0.97} & 0.77 & 0.67 & 0.71 & 0.93 & \textbf{0.97} & 0.85 & 0.91 & 0.98 & 0.95 & 0.95 & 0.97 & \textbf{0.89} \\
 & Bal. Acc. & \textbf{0.90} & 0.74 & 0.64 & 0.78 & 0.91 & \textbf{0.98} & 0.85 & 0.80 & 0.96 & 0.75 & 0.86 & 0.96 & \textbf{0.84} \\
\cline{1-15}
\multirow[c]{6}{*}{AE [256, 128]} & AUC & \textbf{0.92} & 0.82 & 0.57 & \textbf{0.92} & 0.87 & 1.00 & 0.93 & \textbf{0.98} & 0.94 & 0.89 & 0.85 & \textbf{0.96} & 0.89 \\
 & Precision & 0.57 & 0.62 & 0.73 & \textbf{0.62} & 1.00 & 0.67 & 0.63 & 0.16 & 0.48 & 1.00 & 1.00 & 0.41 & 0.66 \\
 & Recall & 0.67 & \textbf{1.00} & 0.95 & 1.00 & 0.82 & 1.00 & \textbf{0.92} & \textbf{1.00} & 0.93 & 0.50 & 0.71 & 0.76 & \textbf{0.86} \\
 & F1 & 0.62 & 0.76 & 0.83 & \textbf{0.77} & 0.90 & 0.80 & 0.75 & 0.27 & 0.63 & 0.67 & 0.83 & 0.53 & 0.70 \\
 & Acc. & 0.96 & 0.62 & 0.70 & \textbf{0.79} & 0.93 & 0.94 & 0.70 & 0.64 & 0.66 & 0.95 & 0.95 & 0.97 & 0.82 \\
 & Bal. Acc. & 0.82 & 0.50 & 0.47 & \textbf{0.83} & 0.91 & 0.96 & 0.71 & 0.80 & 0.73 & 0.75 & 0.86 & 0.87 & 0.77 \\
\bottomrule
\end{tabular}
\end{table*}

\end{document}